\documentclass[aps,twocolumn,amsmath,showpacs,amssymb,superscriptaddress,nofootinbib]{revtex4-1}

\usepackage{graphics,graphicx}
\usepackage{amsmath, amssymb}
\usepackage{dcolumn}% Align table columns on decimal pointA novel scaling relation for $R_A$ is also presentedintroducti
\usepackage{bm}% bold math

\usepackage{hyperref}

\usepackage{color}
\usepackage{amsfonts}

\usepackage[normalem]{ulem}  % \sout{old text} for strikeout

\usepackage{ulem}
\usepackage{color}

\renewcommand\sout{\bgroup \color{blue} \ULdepth=-.5ex \ULset}
\newcommand{\gton}{\mathrel{\lower.9ex \hbox{$\stackrel{\displaystyle
>}{\sim}$}}}
\newcommand{\lton}{\mathrel{\lower.9ex \hbox{$\stackrel{\displaystyle
<}{\sim}$}}}

\begin{document}

\title{Polyakov loop fluctuations in the presence of external fields}

\author{Pok Man Lo}
\affiliation{Institute of Theoretical Physics, University of Wroclaw,
PL-50204 Wroc\l aw, Poland}
\affiliation{Extreme Matter Institute EMMI, GSI,
Planckstr. 1, D-64291 Darmstadt, Germany}
\author{Micha\l {} Szyma\'nski}
\affiliation{Institute of Theoretical Physics, University of Wroclaw,
PL-50204 Wroc\l aw, Poland}
\author{Krzysztof Redlich}
\affiliation{Institute of Theoretical Physics, University of Wroclaw,
PL-50204 Wroc\l aw, Poland}
\affiliation{Extreme Matter Institute EMMI, GSI,
Planckstr. 1, D-64291 Darmstadt, Germany}
\author{Chihiro Sasaki}
\affiliation{Institute of Theoretical Physics, University of Wroclaw,
PL-50204 Wroc\l aw, Poland}

\begin{abstract}

  We study the implications of the spontaneous and explicit Z(3) center symmetry breaking for the Polyakov loop susceptibilities. 
  To this end, 
  ratios of the susceptibilities of the real and imaginary parts, as well as of the modulus of the Polyakov loop
  are computed within an effective model using a color group integration scheme.
  We show that the essential features of the lattice QCD results of these ratios can be successfully captured by the effective approach.
  Furthermore we discuss a novel scaling relation in one of these ratios involving the explicit breaking field, volume, and temperature.

\end{abstract}

%\pacs{12.38.Mh, 11.10.Wx, 25.75.Nq, 11.15.Ha, 24.60.-k, 05.70.Jk}

\maketitle

\section{introduction}

Understanding deconfinement and chiral symmetry restoration, 
as well as exploring their far-reaching consequences~\cite{Braun-Munzinger:2015hba,cs,suganuma}
remain challenging in the study of heavy ion collisions.
A robust description of the phenomenon is necessary for reliably analyzing many observables of experimental interests,
such as fluctuation observables, transport coefficients or the production rate of photon and dilepton~\cite{Braun-Munzinger:2015hba,Ding:2016qdj,Ding:2016hua}.

Although lattice QCD (LQCD) provides first-principles calculations of many of these quantities~\cite{Ding:2016hua,Allton:2005gk,Ejiri:2005wq,Ding:2015ona,Bazavov:2011nk,Aoki:2009sc,Aoki:2006br},
phenomenological models~\cite{cs,Fukushima:2017csk,poly1,poly2,Herbst:2013ufa,plm1,plm2,plm2-2,Sasaki:2013xfa,Sasaki:2012bi,pis,Sasaki:2006ww,matrix-model,Maelger:2017amh,Stiele:2016cfs,Tawfik:2016edq} remain essential for gaining physical understandings and extending results to large baryon chemical potential.

In the limit of infinitely heavy quarks, deconfinement can be identified with the spontaneous breaking of the Z(3) center symmetry~\cite{Greensite, Swanson-confinement}. The Polyakov loop~\cite{Polyakov, tHooft, Yaffe, McLerran} serves as an order parameter for the phase transition.
In some effective models~\cite{plm1,plm2,plm2-2,Sasaki:2013xfa,Sasaki:2012bi} a potential is constructed to describe its behavior.
It is possible to constrain the parameters of the potential using the LQCD results on the thermodynamic pressure~\cite{Borsanyi,Boyd}
and the Polyakov loop~\cite{PL-renorm} in a pure gauge theory. In particular, the latter dictates the locations of the minimum of the potential at different temperatures.

There is another class of independent observables which are sensitive to the Z(3) center symmetry -- the susceptibilities
of the real and imaginary parts, as well as of the modulus of the Polyakov loop
~\cite{pml:su3}.
These quantities measure the fluctuations of the order parameter field.
To describe them in an effective model,
not only the location, but also the curvatures around the minimum of the Polyakov loop potential have to be adjusted~\cite{pml:ug}.

In a pure gauge theory, 
ratios of these susceptibilities have been demonstrated~\cite{pml:su3}
to be excellent  probes of deconfinement.
They display a $\theta$-function like behavior across the transition temperature $T_d$,
with well defined low temperature limits deducible from general theoretical constraints and the Z(3) symmetry.

In a recent study~\cite{lqcd1} these ratios have been computed in numerical simulations of LQCD with 2+1 light flavors.
Unfortunately, the task of extracting useful information from these quantities is more involved than originally thought.
Most importantly, many pertinent features of the ratios are smoothed out in the presence of 
dynamical fermions, as well as after prescribing a renormalization.
In addition, the results are still marred by issues of renormalization scheme dependence and
it is far from clear how to connect them to calculations made in an effective model.

Despite these difficulties, we stress that there are strong theoretical motivations for studying and understanding these ratios.
For one thing, the widely used order parameter, i.e. the renormalized Polyakov loop computed by LQCD,
is a renormalization scheme dependent quantity~\cite{PL-renorm,lqcd2}.
This calls into question the physical relevance of the deconfinement features deduced from it, for example,
the transition temperature $T_d$~\cite{Aoki:2009sc,Aoki:2006br} extracted from its inflection point.
It is therefore crucial to study the deconfinement phenomenon from the perspective of these additional observables,
and investigate whether a coherent picture can be obtained.
They can also be used to signal the strength of the explicit symmetry breaking field.

In this paper we compute the susceptibility ratios within an effective model.
This allows a transparent study of how aspects of center symmetry breaking, explicit and spontaneous, manifests in the ratio observables.
The approach also provides some simple explanations to many features of the LQCD results.

The article is organized as follows: In Sec.~\ref{sec:2} we review the derivation of the Polyakov loop susceptibilities using the color group integration approach.
The method is illustrated by computing one of the ratios, $R_A$, in the presence of explicit symmetry breaking field for a Gaussian model.
In Sec.~\ref{sec:3} we present the effective Polyakov loop potential for this study and analyze the explicit breaking field and volume dependence of the model susceptibility ratios.
A novel scaling relation for $R_A$ is also presented.
In Sec.~\ref{sec:4}, we compare the model results with LQCD calculations.
In Sec.~\ref{sec:5} we present the conclusion.

\section{color group integration}
\label{sec:2}

\subsection{formalism}

In this work we compute the various susceptibility observables using a color group integration scheme~\cite{pml:ug,Miller:1987dg,color_gp}.
The partition function in this approach is expressed as

\begin{align}
  \label{eq:color_gp}
  Z = \int dx dy \, e^{ - V T^3 \, U[x,y] },
\end{align}

\noindent where $(x,y)$ stands for the real and imaginary~\footnote{In this study we shall stay exclusively in the real sector and there is no ambiguity in identifying the longitudinal and transverse directions with the real and imaginary axes, respectively.} part of the Polyakov loop, promoted to a (homogeneous) classical field degree of freedom.
The actual Polyakov loop potential $(U[x,y])$ of choice will be presented in Sec.~\ref{sec:3}.

Expectation value of an arbitrary operator within this approach is computed via

\begin{align}
  \label{eq:gp_av}
  \langle \hat{\mathcal{O}} \rangle = \frac{1}{Z} \, \int dx dy \, \mathcal{O}(x,y) \, e^{ - V T^3 \, U[x,y] }.
\end{align}

\noindent Thus, e.g., the expectation value of the Polyakov loop can be readily obtained from

\begin{align}
 \langle \ell \rangle =  \langle x \rangle + i \, \langle y \rangle.
\end{align}

\noindent Staying within the real sector and considering explicit symmetry breaking along the real axis imply $\langle y \rangle = 0$.
However, fluctuations of the order parameter can be explored along the longitudinal (real) and transverse (imaginary) directions, as well as that of its absolute value:

\begin{align}
  \label{eq:sus1}
    T^3 \chi_L &= VT^3 \, \left( \langle x^2  \rangle  - \langle x  \rangle ^2 \right), \\
  \label{eq:sus2}
    T^3 \chi_T &= VT^3 \, \left( \langle y^2  \rangle  - \langle y  \rangle ^2 \right), \\
  \label{eq:sus3}
    T^3 \chi_A &= VT^3 \, \left( \langle (x^2 + y^2)  \rangle  - \langle \sqrt{x^2+y^2}  \rangle ^2 \right).
\end{align}

\noindent From these,  two independent ratios are derived:

\begin{align}
  R_A = \chi_A/\chi_L \\
  R_T = \chi_T/\chi_L.
\end{align}

\noindent Note that $\chi_A \neq \chi_L + \chi_T$ and $R_A \neq 1 + R_T$.

In the large volume limit, it can be shown that the two susceptibilities in Eqs.~\eqref{eq:sus1}-\eqref{eq:sus2} approach the mean-field results

\begin{align}
  \begin{split}
    T^3 \chi_L &\rightarrow \left({\mathcal{C}}^{-1}\right)_{11} \\
    T^3 \chi_T &\rightarrow \left({\mathcal{C}}^{-1}\right)_{22}
  \end{split}
  ,
\end{align}

\noindent where $\mathcal{C}$ is the correlation matrix~\cite{pml:ug,pml:hq}, defined as

\begin{align}
    \mathcal{C} =
\begin{pmatrix}
  \frac{\partial^2 U}{\partial x \, \partial x}  &   \frac{\partial^2 U}{\partial x \, \partial y} \\
  \frac{\partial^2 U}{\partial y \, \partial x}  &   \frac{\partial^2 U}{\partial y \, \partial y}
  \end{pmatrix}.
\end{align}

\noindent This gives a transparent interpretation of the susceptibilities as the inverse of curvatures of the effective Polyakov loop potential.
Note that all of these quantities are to be evaluated at $(x,y) \rightarrow (x_0, y_0)$, determined from the gap equations

\begin{align}
    \frac{\partial U[x,y]}{\partial x} = 0 =  \frac{\partial U[x,y]}{\partial y}.
\end{align}

\noindent On the other hand, $\chi_A$ in Eq.~\eqref{eq:sus3} does not have a valid mean-field limit.
Nevertheless,  it can be readily computed in the current color group integration scheme.

\subsection{Gaussian model with an explicit symmetry breaking field}

The Gaussian model has proved useful for understanding the low temperature behavior of the susceptibility ratios in a pure gauge system.
Inserting a potential of the form

\begin{align}
U_0 = \alpha \, (x^2 + y^2)
\end{align}

\noindent in Eq.~\eqref{eq:color_gp}, the following non-trivial manifestations of the Gaussian limit can be derived \cite{pml:su3}:

\begin{align}
  \begin{split}
  \label{eq:su3}
    R_A &= 2 - \pi/2 \approx 0.43  \\
    R_T &= 1.
  \end{split}
\end{align}

\noindent These low temperature relations have been verified by lattice calculations in a pure gauge theory.

\begin{figure*}[!ht]
\includegraphics[width=3.355in]{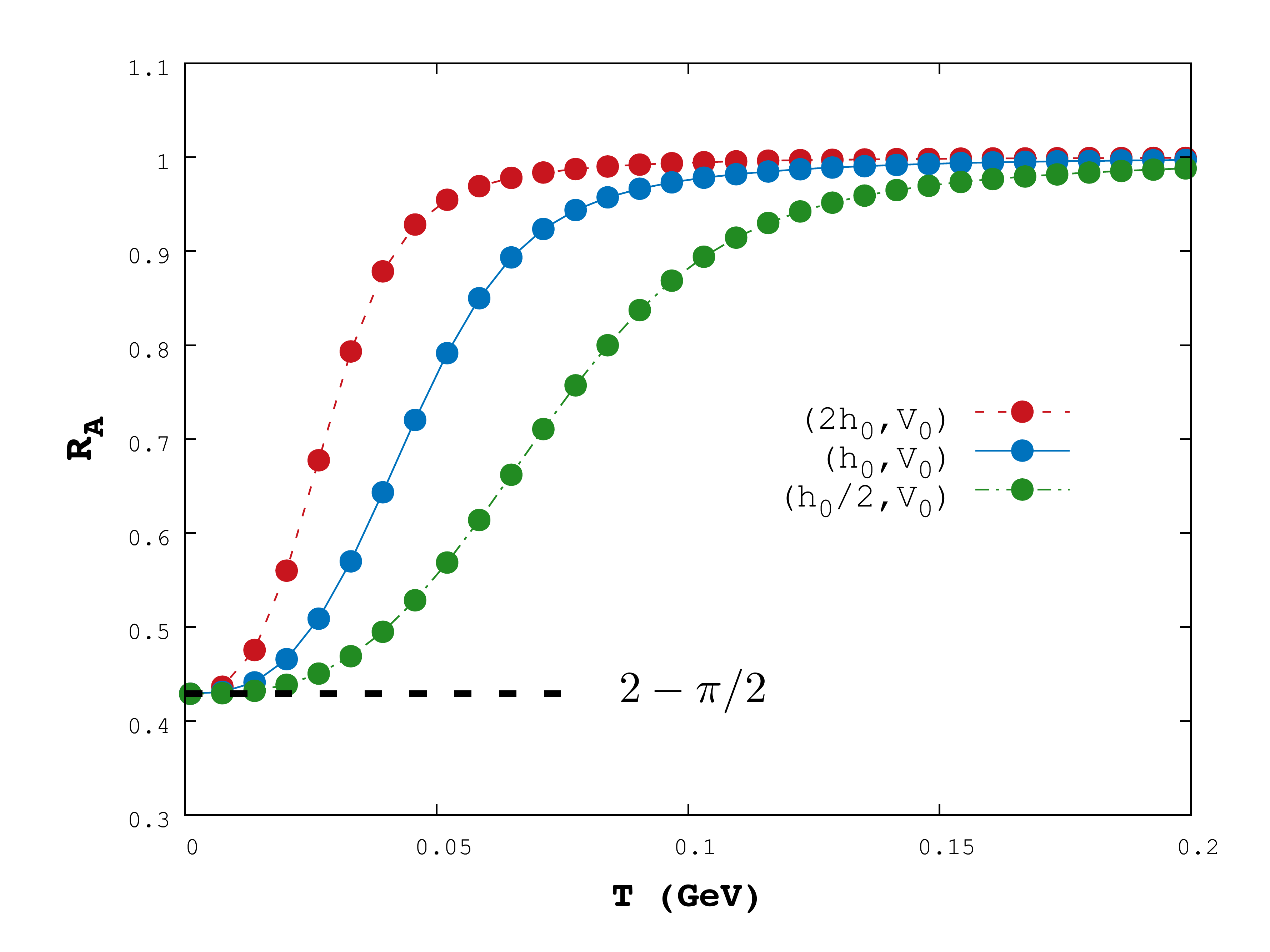}
\includegraphics[width=3.355in]{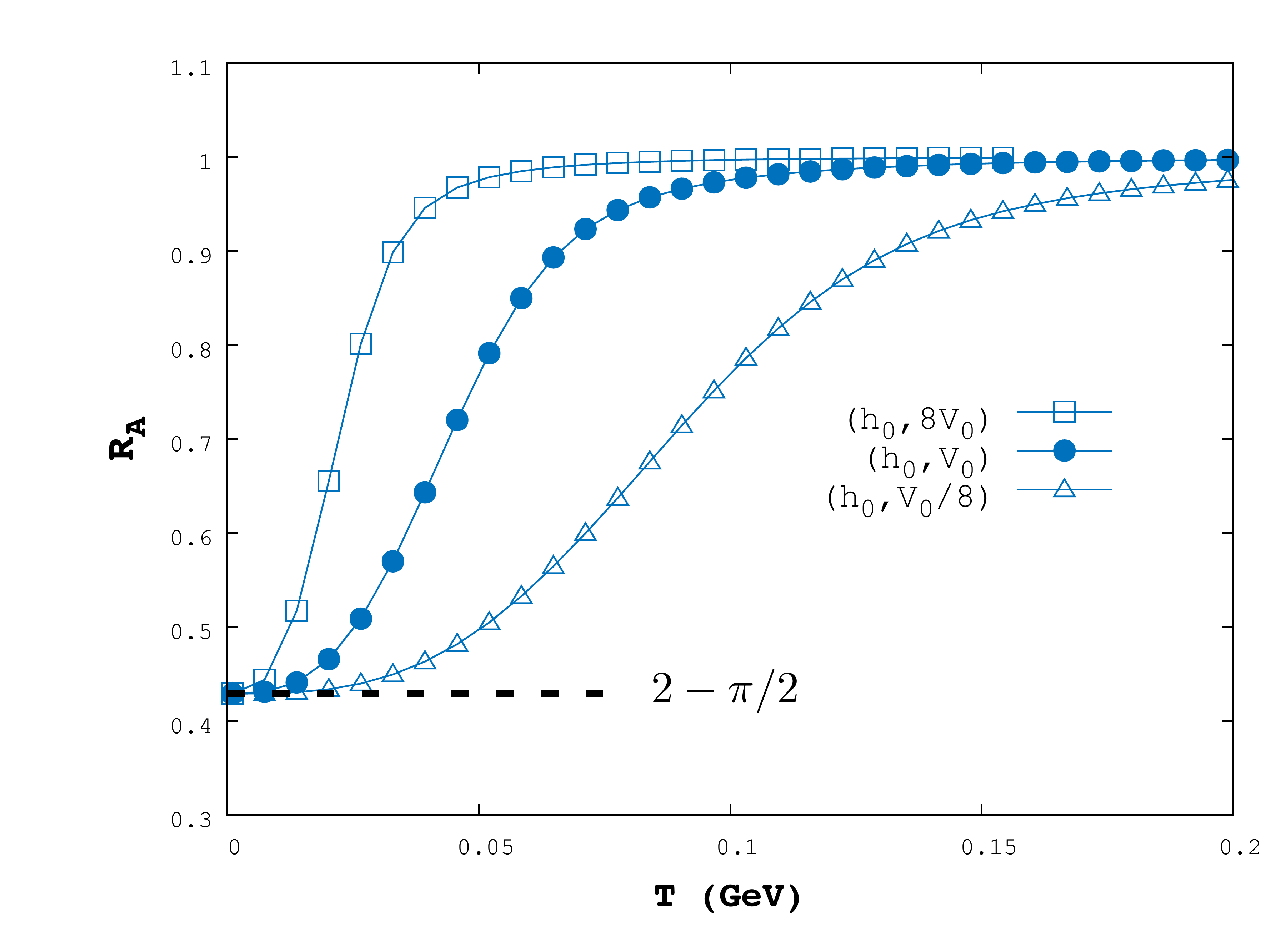}
  \caption{
    The ratio $R_A$ in the Gaussian model as a function of temperature at fixed volume (left) and at fixed explicit symmetry breaking strength (right).
    We adopt the following scheme in presenting our results: different colors (and line types) correspond to different explicit breaking strengths;
    while different symbols denote different volumes. In this numerical study, we fix $\alpha=1$, $h_0 = 1$, and $V_0 =(6.9 \, {\rm fm})^3$.
    }
\label{fig:fig1}
\end{figure*}

To investigate how these ratios behave in QCD with dynamical quarks, we extend the discussion to include finite explicit symmetry breaking.
To this end, we perform a substitution

\begin{align}
    U_0 \rightarrow U_0 - h x.
\end{align}

\noindent It is straightforward to derive an exact expression for $R_A$.~\footnote{A more general case of a double-Gaussian model is considered in the appendix~\ref{appendix}.}. The result reads,

\begin{align}
  \label{eq:RA_gauss}
  R_A &= 2 + 2 \, \xi^2 - \frac{\pi}{2} \, e^{-\xi^2} \times {\mathcal{F}}^2,
\end{align}

\noindent with

\begin{align}
    \mathcal{F} = (1 + \xi^2) \times I_0[\xi^2/2] + \xi^2 \times I_1[\xi^2/2],
\end{align}

\noindent where $I_n(x)$ is the modified Bessel function of the first kind of the n-th order, and

\begin{align}
   \label{eq:xi}
   \xi &= h \times \frac{\sqrt{VT^3}}{2 \sqrt{\alpha}}.
\end{align}

\noindent Especially we extract the following important limits:

\begin{align}
    R_A =
\begin{cases}
  (2 - \pi/2) \times ( 1 + \xi^2 ) & \xi \ll 1 \\
  1 - \frac{1}{4 \xi^2} & \xi \gg 1 \\
\end{cases}
.
\end{align}

\noindent This generalizes Eq.~\eqref{eq:su3} to the case of a finite explicit symmetry breaking.

To gain some familiarity with the ratio $R_A$,
we examine the quantity as a function of temperature:
(1) at fixed volume but for different $h$, and (2) at fixed $h$ but for different values of the volume.
In this numerical study, we fix $\alpha=1$, $h_0 = 1$, and $V_0 =(6.9 \, {\rm fm})^3$.
The results are shown in Fig.~\ref{fig:fig1}.
As expected, the $R_A$ ratio interpolates between the two known limits~\cite{pml:su3}:
from the Z(3)-symmetric phase $(R_A = 2-\pi/2)$ to the Z(3)-broken phase $(R_A = 1)$.
At a fixed volume, increasing the breaking strength $h$ makes the ratio approach unity at lower temperature.
The quantity also exhibits a strong volume dependence, as seen in Fig.~\ref{fig:fig1} right.

All the results presented in Fig.~\ref{fig:fig1} originate from the single expression Eq.~\eqref{eq:RA_gauss}.
In the Gaussian model, the breaking field $h$ always enters via a combination of volume $V$ and the parameter $\alpha$ dictated by Eq.~\eqref{eq:xi}.
This leads to a difficult situation that as $V \rightarrow \infty $, $R_A \rightarrow 1$, regardless of the value of $h$ and temperatures.
To obtain useful information from this quantity, it is necessary to work in a finite volume setting.
Alternatively,  one can study $R_A$ as function of the scaling variable $\xi$.
We shall revisit some of these issues for the full effective model in Sec.~\ref{sec:3}.

\begin{figure*}[!ht]
\includegraphics[width=3.355in]{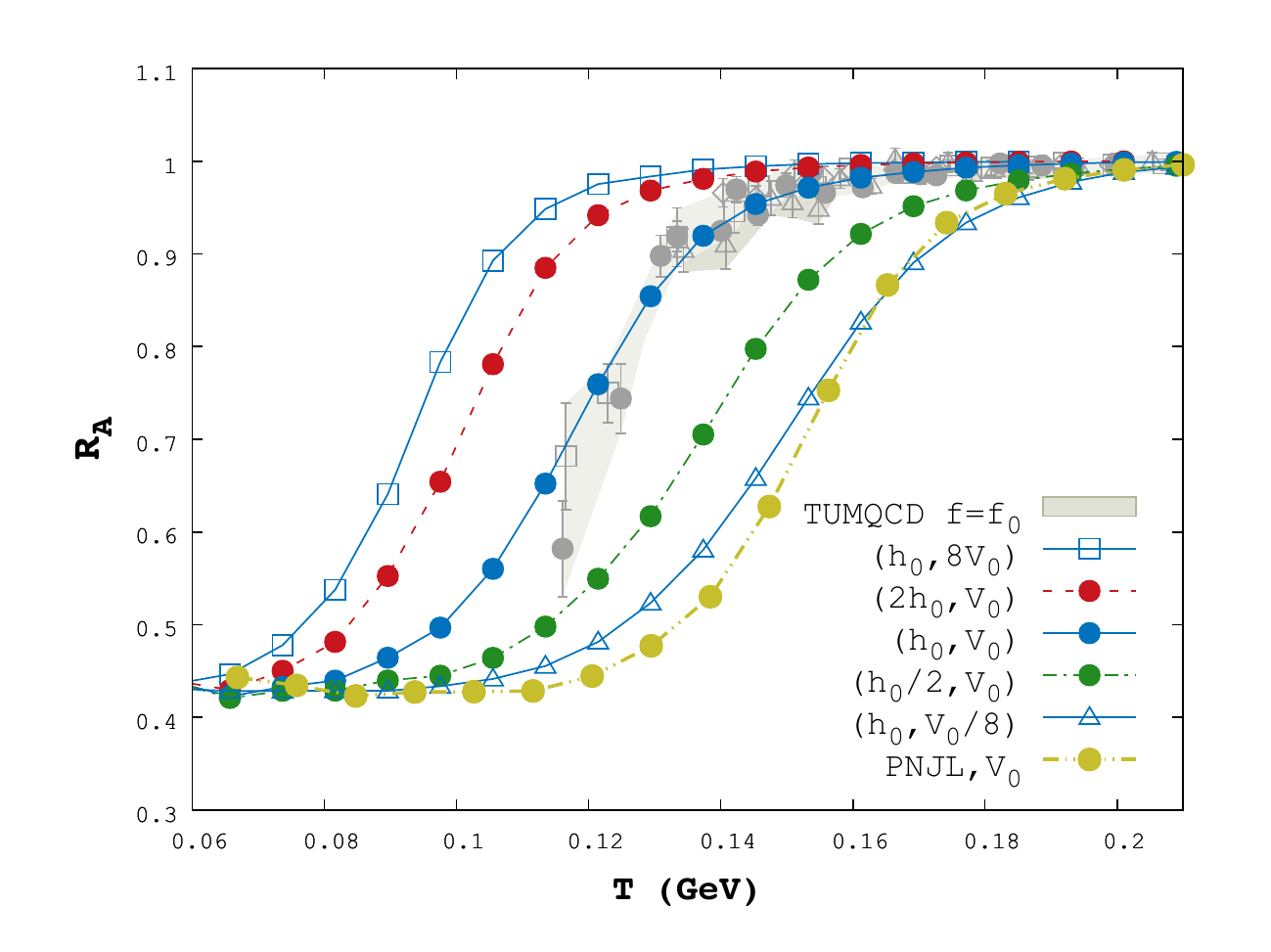}
\includegraphics[width=3.355in]{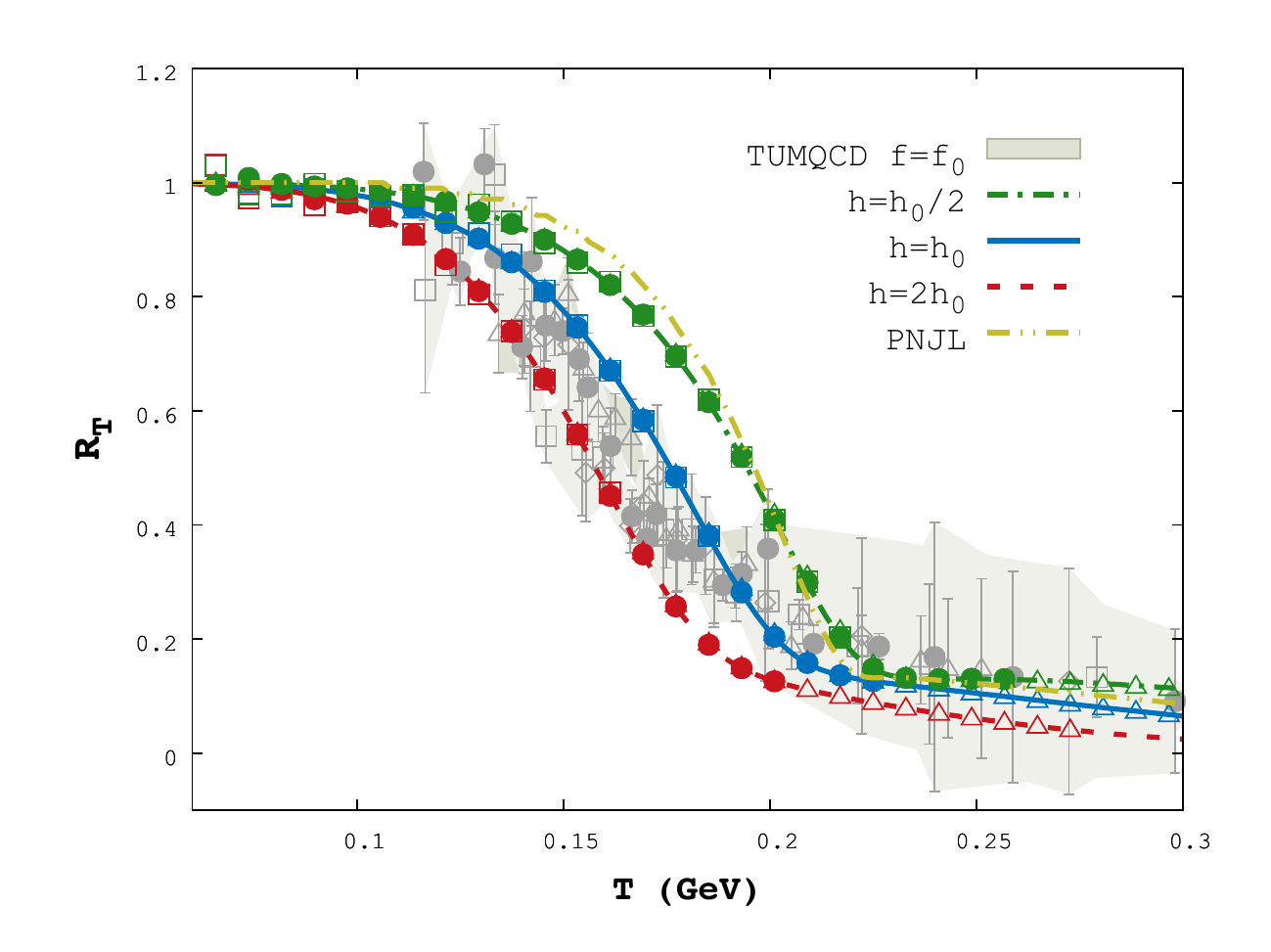}
  \caption{
    The Polyakov loop susceptibility ratio $R_A$ (left) and $R_T$ (right) for the full model Eq.~\eqref{eq:pot}.
    We adopt the same presentation scheme as in Fig.~\ref{fig:fig1} for the results.
    The volume independence of $R_T$ is evident: results with different symbols (but same color) all fall on the same line.
    In this study, $h_0$ is given in Eq.~\eqref{eq:h1}, and $V_0 =(6.9 \, {\rm fm})^3$.
    The ``PNJL'' line denotes the result of a PNJL model for 2+1 light flavors described in Sec.~\ref{sec:4}.
    Also shown are the LQCD results at flow time $f=f_0$ extracted from Ref.~\cite{lqcd1}.
  }
\label{fig:fig2}
\end{figure*}

\section{Polyakov loop susceptibility ratios within an effective model}
\label{sec:3}

\subsection{effective Polyakov loop potential}

The Gaussian model discussed above, though generalized to include an explicit symmetry breaking field,
does not describe the spontaneous Z(3) symmetry breaking.~\footnote{The Gaussian model gives $R_T=1$ for all temperatures.}
To examine the susceptibility ratios in a setting that is relevant to QCD,
a Polyakov loop potential~\cite{pml:ug},  capable of handling the latter aspect  will be employed:

\begin{align}
\label{eq:pot}
\begin{split}
  U_G &= -\frac{A}{2} \times \bar{\ell} \ell + B \times \ln M_H(\ell,\bar{\ell})\\
  &+\frac{C}{2} \times (\ell^3+{\bar{\ell}}^3) + D \times(\bar{\ell} \ell)^2.
\end{split}
\end{align}

\noindent Here $M_H(l,\bar{l})$ is the SU(3) Haar measure

\begin{align}
  M_H(l,\bar{l}) = 1 - 6 \, \bar{\ell}\ell + 4 \, (\ell^3+{\bar{\ell}}^3) - 3 \, (\bar{\ell}\ell)^2,
\end{align}

\noindent The temperature dependent model parameters $(A, B, C, D)$ are given in Ref.~\cite{pml:ug} and will not be repeated here.
Note that to implement the color group integration scheme in Eq.~\eqref{eq:color_gp}, we use

\begin{align}
\begin{split}
  {\ell} &= x + i y \\
  \bar{\ell} &= x - i y.
\end{split}
\end{align}

The potential $U_G$ in Eq.~\eqref{eq:pot} is particularly suited for the current study.
Most importantly, the known susceptibilities at zero explicit breaking are reproduced by construction.
This is not the case for other commonly used Polyakov loop potentials~\cite{plm1,plm2,plm2-2}.
For example, the polynomial potential introduced in Ref.~\cite{plm1} leads to the result $R_T > 1$ for $ T > T_c$,
which is another manifestation of the ``negative susceptibility'' problem discussed in Ref.~\cite{poly2,Sasaki:2006ww}.
Imposing the Haar measure to the potential~\cite{plm2, plm2-2} effectively restricts the Polyakov loop to the target region
and thus improves the theoretical description. In fact, the present model builds on this observation
and further constrains the curvatures of the potential using the available LQCD results~\cite{pml:su3}
on the susceptibilities in a pure gauge theory.

\begin{figure*}[!ht]
\includegraphics[width=3.355in]{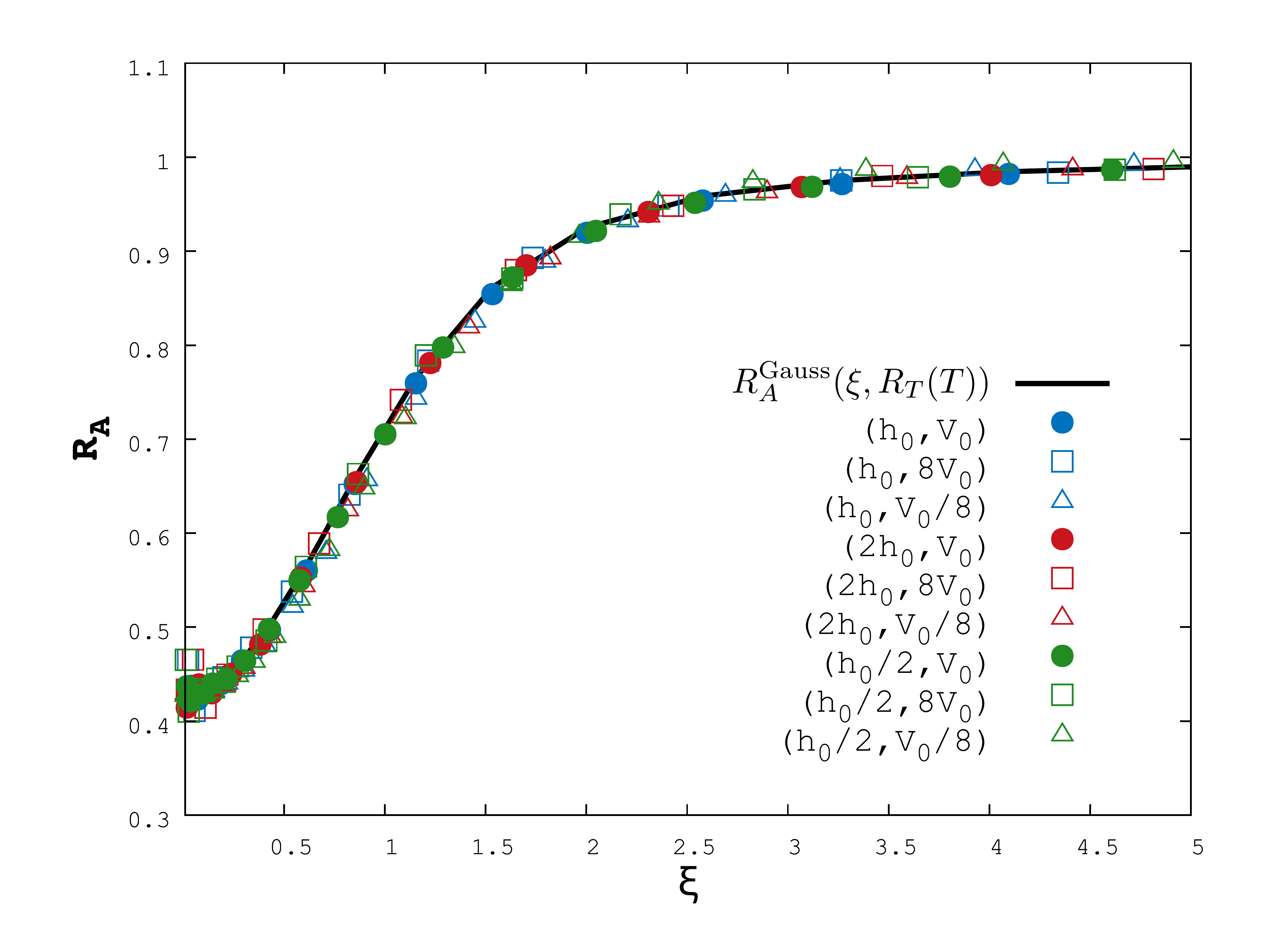}
\includegraphics[width=3.355in]{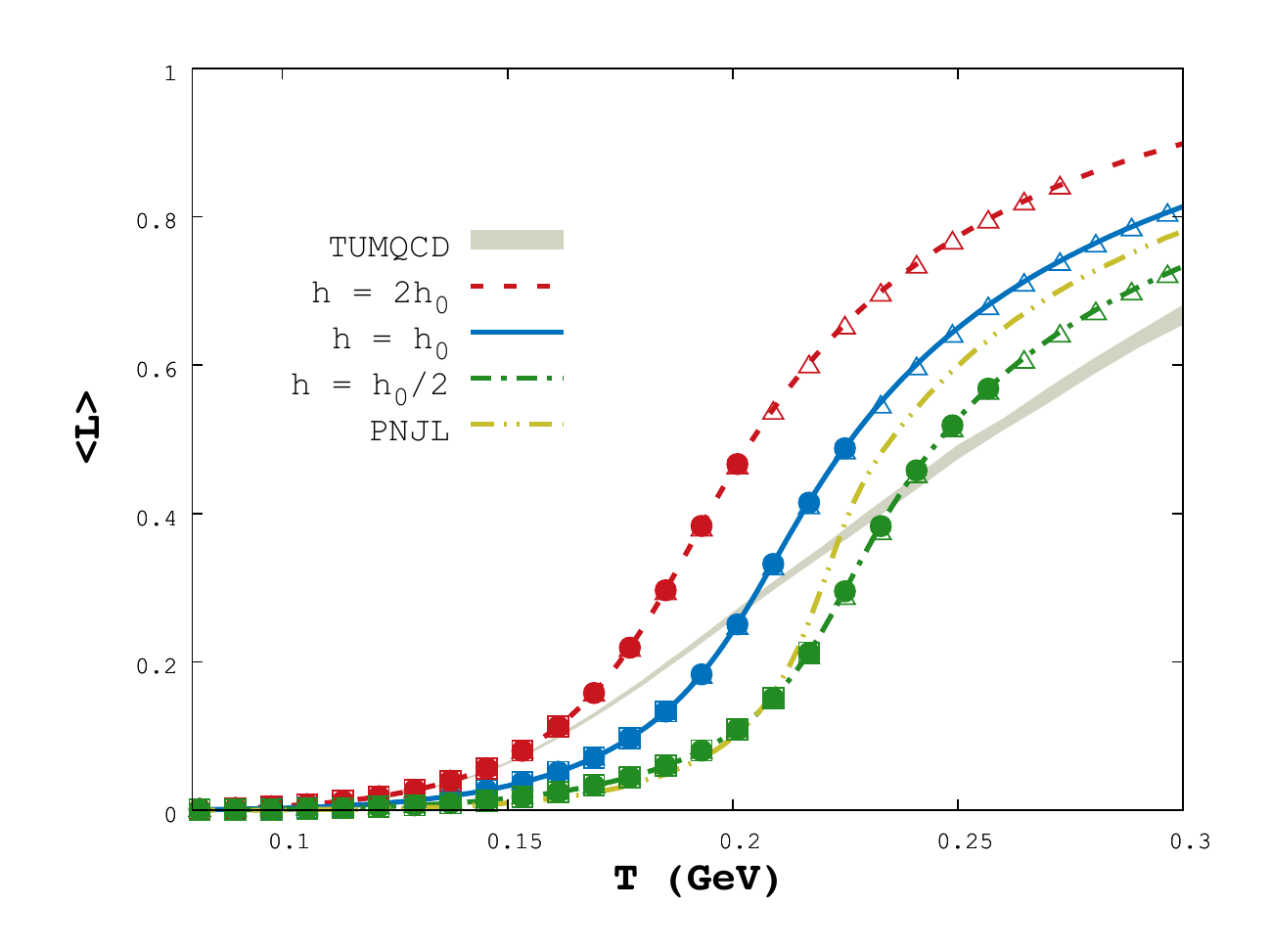}
  \caption{
    Left: The $R_A$ ratio for the full model plotted as a function of the scaling variable $\xi$ (Eq.~\eqref{eq:xi2}).
  Model results at various $(h,V)$ collapse into a single line determined by the generalized Gaussian formula in Eq.~\eqref{eq:exact_RA}.
  Right: Results of the Polyakov loop computed in the model and the LQCD results from Ref~\cite{lqcd1}. 
  The ``PNJL'' line denotes the result of a PNJL model for 2+1 light flavors described in Sec.~\ref{sec:4}.
  }
\label{fig:fig3}
\end{figure*}

Furthermore, we consider a linear explicit breaking term $ h \times x $.
The functional form of the breaking field $h$ is known within the approximation of a one-loop expansion of the fermionic determinant~\cite{pml:hq,PhysRevD.85.114029}. 
As we aim at understanding the ratios on the qualitative level, we shall employ the following basic form for $h=c \times h_0$,  with

\begin{align}
  \label{eq:h1}
  h_0 &= 2 \, h_q(T, m_l) + h_q(T, m_s)
\end{align}

\noindent where

 \begin{align}
 \label{eq:h2}
 \begin{split}
 h_q(T,m) &= \frac{6}{\pi^2 T^3} \, \int_0^\infty dk \, k^2 \, \left ( \frac{e^{-E(k)/T} + e^{-2 E(k)/T}}{1 + e^{-3 E(k)/T}} \right) \\
   \\
 E(k) &= \sqrt{k^2 + m^2}.
 \end{split}
 \end{align}

\noindent For quark masses we choose $m_l = 5$ MeV for up and down quarks and $m_s = 100$ MeV for the strange quarks.~\footnote{
The case of a (2+1)-PNJL model is addressed in Sec.~\ref{sec:4}.}
Eq.~\eqref{eq:h2} improves on the previous study~\cite{pml:hq} by taking the quantum statistics of quarks into account.
In practice this leads to an increase of $\approx 10 \%$ in the total strength $h_0$ over the one imposing a further Boltzmann approximation.
We shall also allow for an arbitrary prefactor $c$ to manipulate the strength of $h$.
In fact, an attempt will be made to infer its magnitude from the LQCD results on the ratio observables.

In a previous study~\cite{pml:hq} we have calculated the critical strength of the breaking field, $h_{\rm crit.}$, for the phase transition
to turn from the first order to the second order, that is, the critical end point (CEP) for the Z(3) transition:
~\footnote{We correct a typographical mistake in Ref.~\cite{pml:hq}.}

\begin{align}
  h_{\rm crit.} \approx 0.144.
\end{align}

\noindent The breaking field $h_0$ in Eq.~\eqref{eq:h1} exceeds this limit for all temperatures of interest, meaning that a crossover transition is expected.
This is evident in the ratio observables computed in the full model, as shown in Fig.~\ref{fig:fig2}.

Starting with the ratio $R_A$, shown in Fig.~\ref{fig:fig2} \, left,
we first notice the similarity between the full model results and those from the Gaussian model.
Indeed, the $R_A$ ratio interpolates between the two known theoretical limits: $(2-\pi/2\approx 0.43)$ and 1.
The expected behaviors from varying the breaking strength and the volume are also verified.

Turning now to the ratio $R_T$, shown in  Fig.~\ref{fig:fig2} \, right, the immediate observation is the volume independence of the quantity.
This is evident from the fact that results with different symbols (but same color) all fall on the same line.
This suggests that the finite volume $V_0 =(6.9 \, {\rm fm})^3$ we selected is sufficiently large.
Indeed, we have checked that the ratio $R_T$ approaches the mean field value, dependent only on the intensive
variables $T$ and $h$.
Increasing the breaking strength $h$ makes the ratio deviates from the known Z(3)-symmetric limit $(R_T=1)$ at lower temperature.
On the other hand, the value of $R_T$ at large temperatures is not dictated by the Z(3) symmetry.
Instead, it can be related to the color screening properties of the QCD medium.

Before proceeding to compare the effective model calculations with LQCD (see Sec.~\ref{sec:4}),
we first discuss an interesting observation of $R_A$ ratio, namely, a scaling relation inspired by the Gaussian model.

\subsection{scaling relation of $R_A$}

The Polyakov loop potential $U_G$ employed in Eq.~\eqref{eq:pot} is clearly non-Gaussian.
Nevertheless, we can consider a generalized double-Gaussian approximation to the potential:

\begin{align}
  U_1 = \alpha_1 \, x^2 + \alpha_2 \, y^2 - \tilde{h} x,
\end{align}

\noindent where the model parameters $\alpha_1$, $\alpha_2$, and $\tilde{h}$ are constructed to match

\begin{align}
  \label{eq:gmatch}
  \begin{split}
    \alpha_1 &= \frac{1}{2 \, T^3 \, {\chi^{(0)}_L}} \\
    \alpha_2 &= \frac{1}{2 \, T^3 \, {\chi^{(0)}_T}} \\
    \tilde{h} &= h + \frac{ {\langle \ell \rangle}^{(0)}}{ T^3 \, {\chi^{(0)}_L }},
  \end{split}
\end{align}

\noindent with

\begin{align}
  \begin{split}
    \chi^{(0)}_L &=  {\chi_L}(T, h = 0, V \rightarrow \infty) \\
    \chi^{(0)}_T &=  {\chi_T}(T, h = 0, V \rightarrow \infty) \\
    {\langle \ell \rangle}^{(0)} &= {\langle \ell \rangle}(T, h = 0, V \rightarrow \infty).
  \end{split}
\end{align}

\noindent These coefficients can be readily obtained by a mean-field calculation of the original potential at vanishing breaking field.

It is clear that the approximation scheme operates by constructing local double-Gaussian potential along the line of minima of $U_G$.
The advantage of performing such an expansion is that it allows a direct computation of the ratio $R_A$ with an
equation analogous to the single Gaussian limit studied previously in Eq.~\eqref{eq:RA_gauss}. The generalized equation reads

\begin{align}
\label{eq:exact_RA}
  R_A(\xi, R_T) = 1 + R_T + 2 \xi^2 - \frac{2}{\pi} \, R_T \, e^{-2 \xi^2} \left[ \mathcal{F}(\xi,R_T) \right]^2,
\end{align}

\noindent where $\mathcal{F}$ can be obtained with an integral involving the modified Bessel function. (Details in the appendix~\ref{appendix})
According to this equation, the functional dependence in $(T, h, V)$ of $R_A$ can be uniquely determined by the scaling variable $\xi(T, h, V)$ and $R_T(T,h)$, via

\begin{align}
  \label{eq:xi2}
  \begin{split}
    \xi &= \tilde{h} \times \frac{\sqrt{VT^3}}{2 \sqrt{\alpha_1}} \\
    R_T &= \frac{\alpha_1}{\alpha_2}.
\end{split}
\end{align}

\noindent This translates to the following: Provided that the generalized Gaussian approximation is valid, all the data point of $R_A(T,h,V)$ will collapse on a single universal line when plotted against $\xi$, with the choice a ``physical'' $R_T = R_T(T,h)$. A direct numerical computation confirms that it is indeed the case,  and the result is shown in Fig.~\ref{fig:fig3} \, left.

While the observation is theoretically interesting, it also indicates a rather limited information contained in this observable.
For example, the key information about the magnitude of the explicit breaking field can as well be extracted from $R_T$.
Nevertheless, Eq.~\eqref{eq:exact_RA} may serve as a useful diagnostic for analyzing $R_A$.

\section{comparison with LQCD results and a PNJL model}
\label{sec:4}

As discussed in the introduction, the renormalized Polyakov loop computed by LQCD is a renormalization scheme dependent quantity~\cite{lqcd1,lqcd2}.
It obscures the physical relevance of the derived deconfinement features,
e.g. the $T_d$ extracted from the inflection point, and complicates the comparison of LQCD results with those obtained in an effective approach.

One of the original motivations for introducing the susceptibility ratios as probe of deconfinement
is the removal of both the cutoff and the scheme dependence.
The assumption is that if the Polyakov loop susceptibilities are renormalized the same way as the Polyakov loop,
the multiplicative renormalization factor will be canceled against each other.

Contrary to this expectation, recent studies~\cite{lqcd1,lqcd2} report a substantial cutoff dependence in these ratios in QCD with 2+1 light flavors.
This is evident from the $N_\tau$-dependence observed in the ``bare data'' of $R_A$ and $R_T$ in Ref.~\cite{lqcd1}.
This seems to suggest that renormalizing the Polyakov loop alone does not guarantee the renormalization of the susceptibilities.

In the effective model, the behavior of $R_T$ is largely determined by the explicit breaking field $h$.
It is possible, therefore, that the cutoff dependence observed in $R_T$
can be associated with the cutoff dependence of $h$.
In fact, it is non-trivial to obtain a continuum extrapolation of the explicit breaking strength $h$ from LQCD
that is suitable for comparison with effective model~\cite{pml:hq, tina, Fromm:2012eb, Philipsen:2016wjt, Fischer:2014vxa}.

Furthermore, if we mimic the $N_\tau$-dependence of the ratio observables
as a change of prefactor in $h$ in the effective model, we can reproduce the same trend in the
ordering of curves of $R_T$ and $R_A$, namely, from top to bottom for $R_A$ in increasing $N_\tau$ and
the reverse order for $R_T$. (See Fig. 17 and 18 of Ref.~\cite{lqcd1}.)
This suggests that the two sets of ``bare'' data are connected, and the connection may be due to $h$.

\begin{figure}[!ht]
\includegraphics[width=3.355in]{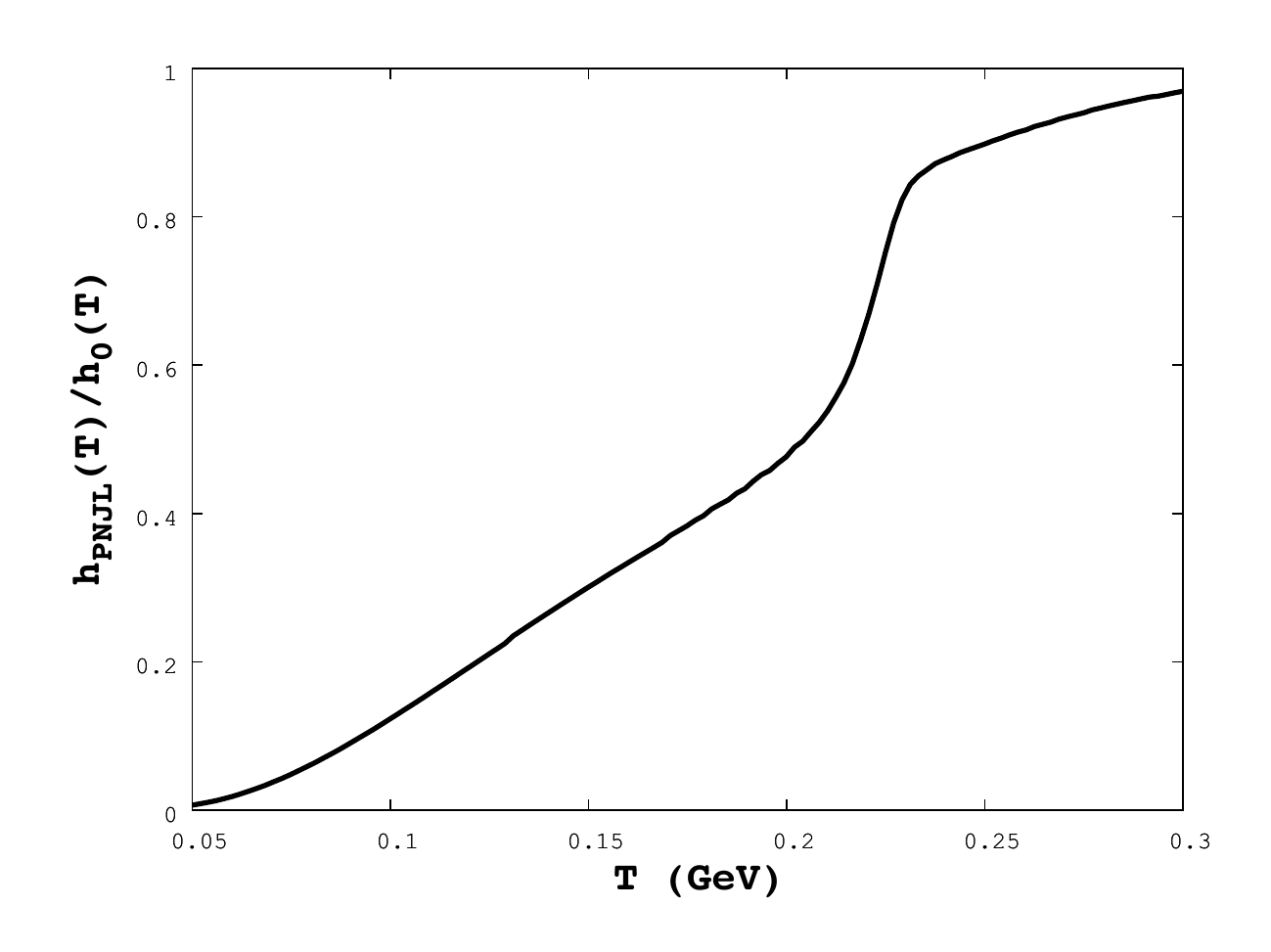}
  \caption{ The ratio of the effective breaking field (Eq.~\eqref{eq:h1}) computed using the constituent quark masses obtained
  in a (2+1)-PNJL model to the same quantity computed with the current quark masses. This ``prefactor'' summarizes
  the effects of chiral symmetry (within the model) on the explicit Z(3) breaking strength.}
\label{fig:fig4}
\end{figure}

Using the gradient flow method~\cite{lqcd1,lqcd2}, it is possible to renormalize the susceptibilities and the ratio observables.
We have selected  LQCD results with the  ``$f = f_0$'' flow time to compare with our effective model calculations.
They are presented in Fig.~\ref{fig:fig2}. We note that a reasonable agreement can be attained
if we choose an explicit breaking field of strength $ \approx (1 - 2) \times h_0$ and a physical volume of $V_0 \approx (6.9 \, {\rm fm})^3$.

It is straightforward to extend this study to incorporate effects from the spontaneous chiral symmetry breaking.
For this purpose, we study a (2+1)-PNJL model, combining the NJL model in Refs.~\cite{Klevansky:1992qe,Hatsuda:1994pi} 
and the Polyakov loop potential $U_G$ in Eq.~\eqref{eq:pot}.
The computation is similar to the one presented in Sec.~\ref{sec:2}, 
and we simply show the major results in Fig.~\ref{fig:fig2} and~\ref{fig:fig3}.
These results can be readily understood by studying an effective Z(3) breaking strength for the PNJL model.
It can be computed via Eq.~\eqref{eq:h1}, except for using the constituent quark masses in lieu of the current ones.
The behavior of such an effective breaking strength compared to $h_0$, i.e. the prefactor $c_{\rm PNJL}(T)$ for the PNJL model, 
is shown in Fig.~\ref{fig:fig4}.
It is evident that the PNJL model leads to a smaller explicit Z(3) breaking, and the observables $R_T$ and $R_A$ behave accordingly.
However, the LQCD results of ratio observables seem to indicate a stronger breaking strength.
This suggests that a naive implementation of the coupling between quarks and the Polyakov loop may be inadequate and
a more sophisticated treatment including the back-reaction of dynamical quarks on the gauge sector,
which could modify the Polyakov loop potential $U_G$, may be necessary.

\begin{figure*}[!ht]
\includegraphics[width=3.355in]{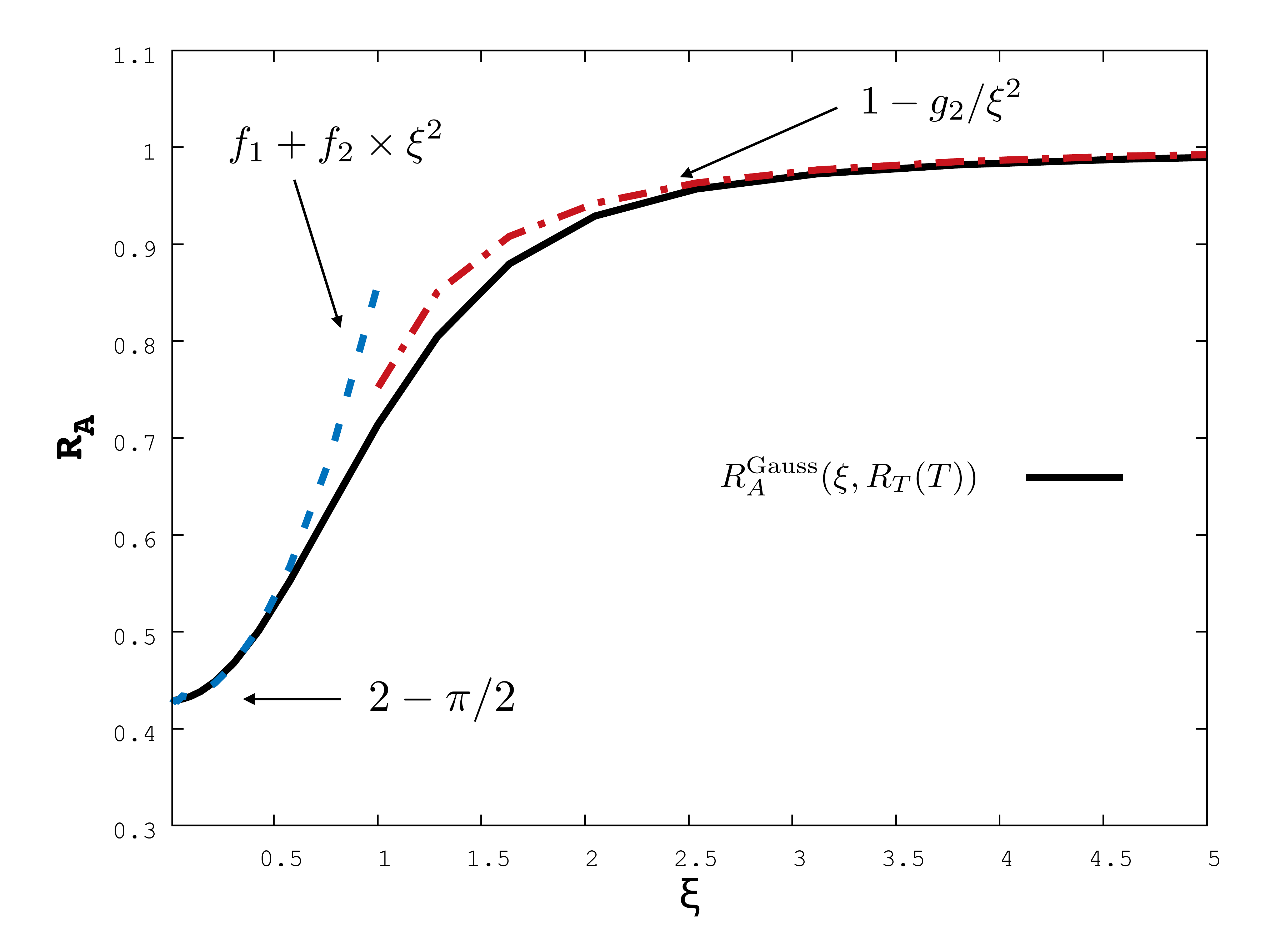}
\includegraphics[width=3.355in]{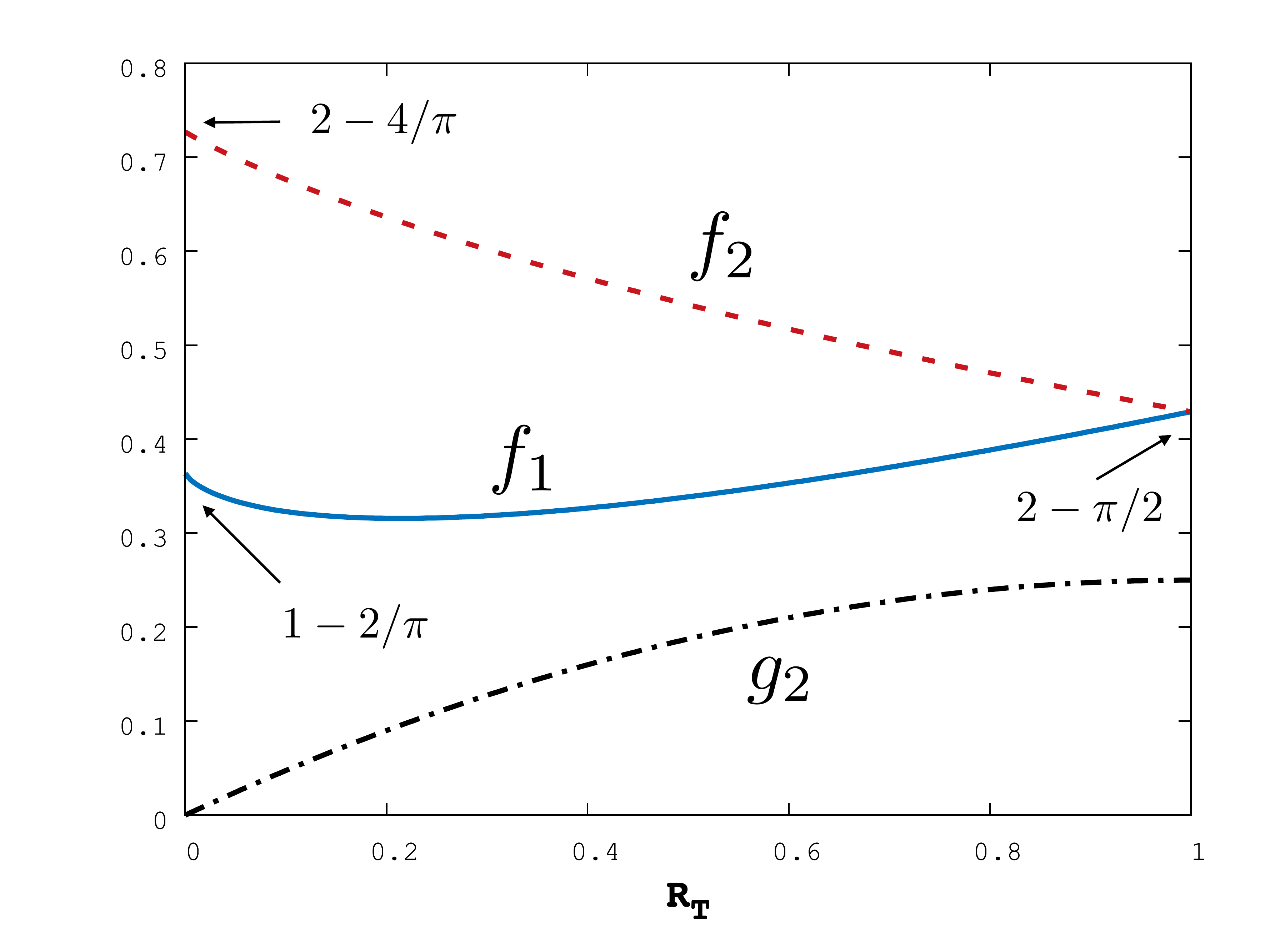}
  \caption{Left: Schematic plot of $R_A(\xi, R_T)$ as a function of $\xi$ for a given $R_T(T)$.
           Right: $R_T$-dependence of various expansion parameters of $R_A(\xi, R_T)$.
  }
\label{fig:fig5}
\end{figure*}

Furthermore, there is still substantial ``flow time'' dependence in these observables,
reflecting a further renormalization prescription dependence.
~\footnote{It was, however, reported~\cite{lqcd2} that the ratio observables exhibit a milder flow time dependence than the susceptibility.}
In particular, the large flow time ($f = 3 f_0$) result of $R_T$ shows a relatively low value $(\approx 0.7)$
at low temperatures, instead of the expected Z(3)-symmetric limit of unity.
It is still possible to describe such $R_T$ within our effective model, though it requires a rather large $h$.
This also naturally explains the observation that $R_A \rightarrow 1$ at large flow time.
We however find this situation unsatisfactory, since it is more natural to expect the physical breaking field $h$
to be free of the renormalization scheme dependence. And it is this quantity that we hope to extract from the LQCD.

For reference we also compute the Polyakov loop in the effective model and compare with the LQCD result.
This is shown in Fig.~\ref{fig:fig3} \, right.
Unlike the case of the ratio observables, we see that the effective model essentially fails to describe the LQCD result.
Similar discrepancy has been reported by the matrix model~\cite{poly1}.
This may again be due to the scheme dependence and we shall explore this topic in more detail in a future publication.

\section{conclusion}
\label{sec:5}

This study has demonstrated how features of deconfinement emerge
in the ratios of Polyakov loop susceptibilities within an effective model.
For the ratio $R_A$, we find a characteristic volume dependence along
with temperature and the explicit Z(3) symmetry breaking strength $h$.
In a Gaussian approximation scheme, all these can be subsumed into a single scaling equation.
For the ratio $R_T$, we find a minimal volume dependence,
which makes it a robust probe of the strength of the explicit breaking term.

On a qualitative level, the effective model is capable of describing many features of the LQCD results.
These include the low and high temperature limits of the ratios,
and the connection between $R_A$ and $R_T$, possibly via $h$.
Nevertheless, it is important to bear in mind that the LQCD results,
which we compare our model results to and estimate the effective strength of $h$ from,
still suffer from a renormalization scheme dependence.
This is an urgent issue to be tackled to achieve a meaningful comparison of effective approaches with LQCD,
and will be pursued in future works.

It would also be interesting to investigate how the susceptibility observables
behave at large baryon densities~\cite{Baym:2017whm}
and react to other external fields, e.g. a magnetic field~\cite{Schafer:2015wja,b-field}.
Since the curvatures dictate how reluctant the system is to deviate from the equilibrium position
in the presence of external disturbances, we expect these susceptibilities would be crucial to successfully describing the system.

\acknowledgments

We thank Bengt Friman for stimulating discussions.
We are also grateful to Frithjof Karsch, Olaf Kaczmarek, Peter Petreczky and Owe Philipsen for the constructive comments.
MS acknowledges partial support from the NCN grant Polonez 2016/21/P/ST2/04035.
This work was partly supported by the Polish National Science Center (NCN) under Maestro Grant No. DEC-2013/10/A/ST2/00106, 
and by the ExtreMe Matter Institute EMMI at the GSI Helmholtzzentrum fuer Schwerionenphysik, Darmstadt, Germany.

\appendix

\section{Exact expression of $R_A(\xi, R_T)$ in a Gaussian model}
\label{appendix}

In this appendix, we present the analytic expression of the ratio $R_A$ in the presence of external field
within a Gaussian model. Starting with the model partition function

\begin{align}
Z_{\rm Gauss} = \int d x d y \, e^{ - \left( \mathcal{A}_1 x^2 + \mathcal{A}_2 y^2 - \mathcal{H} x \right) },
\end{align}

\noindent the ratio $R_A$ can be computed by

\begin{align}
  R_A &= \frac{\langle \left( x^2 + y^2 \right) \rangle - \langle \left( \sqrt{x^2 + y^2} \right) \rangle^2}{\langle x^2  \rangle - \langle x \rangle^2},
\end{align}

\noindent where

\begin{align}
  \langle \mathcal{O} \rangle = \frac{1}{Z_{\rm Gauss}} \, \int d x d y \, \mathcal{O}(x,y) \, e^{ - \left( \mathcal{A}_1 x^2 + \mathcal{A}_2 y^2 - \mathcal{H} x \right) }.
\end{align}

The ratio $R_A$ depends on the model parameters $\left\{ \mathcal{A}_1, \mathcal{A}_2, \mathcal{H} \right\}$ via the following combinations

\begin{align}
  R_A(\mathcal{A}_1, \mathcal{A}_2, \mathcal{H}) \rightarrow R_A(\xi, R_T)
\end{align}

\noindent where

\begin{align}
  \begin{split}
    \xi = \frac{\mathcal{H}}{2 \sqrt{\mathcal{A}_1}} \\
    R_T = \frac{\mathcal{A}_1}{\mathcal{A}_2}.
  \end{split}
\end{align}

\noindent The exact expression reads

\begin{align}
  R_A(\xi, R_T) = 1 + R_T + 2 \xi^2 - \frac{2}{\pi} \, R_T \, e^{-2 \xi^2} \left[ \mathcal{F}(\xi,R_T) \right]^2,
  \label{eq:exact_RA-2}
\end{align}

\noindent with

\begin{align}
  \begin{split}
  \mathcal{F}(\xi, R_T) =& \frac{1}{\sqrt{\pi}} \, \int_{-\infty}^{\infty} d x \, e^{-x^2 + 2 \xi x} \times  \frac{x^2}{2 R_T } \times \\
  & e^{ \frac{x^2}{2 R_T} } \times \left( K_0[\frac{x^2}{2 R_T}] + K_1[\frac{x^2}{2 R_T}] \right),
  \end{split}
\end{align}

\noindent where $K_n$ is the modified Bessel function of the second kind of the n-th order.

It is instructive to study the limits of $\xi \ll 1$ and $\xi \gg 1$ for a general $R_T$:

\

\begin{align}
    R_A \approx
\begin{cases}
  f_1 + f_2 \times  \xi^2 & \, \xi \ll 1 \\
   1 - g_2 \times \frac{1}{\xi^2} & \, \xi \gg 1
\end{cases}
.
\end{align}

\noindent where

\begin{align}
  \begin{split}
  f_1 &= 1 + R_T - \frac{2}{\pi} \left( E[1-R_T] \right)^2 \\
  f_2 &= 2 - \frac{4}{\pi} \frac{E[1-R_T]}{1-R_T} \times \\
  & \left(  E[1-R_T] - R_T \times K[1-R_T] \right) \\
  g_2 &= \frac{ (2-R_T) \, R_T}{4}.
  \end{split}
\end{align}

\noindent Here $K, E$ are the complete elliptic integral of the 1st and 2nd kind respectively, defined as

\begin{align}
  \begin{split}
    K(x) &= \int_0^{\pi/2} d \theta \, (1-x \sin^2 \theta)^{-1/2} \\
    E(x) &= \int_0^{\pi/2} d \theta \, (1-x \sin^2 \theta)^{1/2}.
  \end{split}
\end{align}

\noindent For $R_T = 0$, we obtain the SU(2)~\cite{pml:su3,Engels:1998nv} limit:

\begin{align}
  R_A(\xi,R_T=0) \approx
\begin{cases}
  (1-2/\pi)  \times ( 1 + 2 \, \xi^2 )  & \, \xi \ll 1 \\
    1 - \frac{2}{\sqrt{\pi}} \frac{1}{\xi} e^{-\xi^2}  & \, \xi \gg 1
\end{cases}
.
\end{align}

A schematic plot of $R_A(\xi, R_T)$ in Eq.~\eqref{eq:exact_RA-2}, as a function of $\xi$ for a given $R_T(T)$, is illustrated in Fig.~\ref{fig:fig5} left, together with
the functional dependence on $R_T$ of various expansion coefficients in Fig.~\ref{fig:fig5} right.

\end{document}